\documentclass[12pt]{article}

\input epsf
\def\beq{\begin{eqnarray}}
\def\eeq{\end{eqnarray}}

\def\mpl{M_{\rm Pl}}

\def\l{{\cal L}}

\def\lsim{\mathrel{\rlap{\lower3pt\hbox{\hskip0pt$\sim$}}
     \raise1pt\hbox{$<$}}}         %less than or approx. symbol
\def\gsim{\mathrel{\rlap{\lower4pt\hbox{\hskip1pt$\sim$}}
     \raise1pt\hbox{$>$}}}         %greater than or approx. symbol

\newcommand{\ud}{\mathrm{d}}

\newcommand{\fD}{\mathcal{D}}

\usepackage{amsmath}
\usepackage{amsfonts}

\begin{document}

\begin{flushright}
{ NYU-TH-05/06/01}
\end{flushright}
\vskip 0.9cm

\centerline{\Large \bf Classically Constrained Gauge Fields and Gravity}
\vspace{0.3in}
%\vspace{0.3in} 
\vskip 0.7cm
\centerline{\large Gregory Gabadadze and Yanwen Shang}
\vskip 0.3cm
\centerline{\em Center for Cosmology and Particle Physics}
\centerline{\em Department of Physics, New York University, New York, 
NY, 10003, USA}

\vskip 1.9cm

\begin{abstract}

We study gauge and gravitational field theories in  
which  the gauge fixing conditions 
are imposed as constraints on classical 
fields. Quantization of fluctuations 
can be performed in a BRST invariant manner, 
while the main novelty is that  the classical equations 
of motion  admit solutions that are not present 
in the standard approach. Although the new solutions 
exist for both gauge and gravitational fields, 
one interesting example we consider in detail 
is constrained gravity  endowed with a nonzero 
cosmological constant. This theory, unlike General Relativity, 
admits  two maximally symmetric solutions one of which is a 
flat space, and another one is a curved-space solution of GR.
We argue that, due to  BRST symmetry,  the classical solutions  
obtained in these theories  
are not ruined by  quantum effects. 
We also comment on  massive deformations of the 
constrained models. For both  gauge and gravity fields
we point out that the propagators of the massive quanta 
have  soft ultraviolet behavior and smooth transition 
to the massless limit. However, nonlinear stability may require 
further modifications of the massive theories.

\end{abstract}

\vspace{3cm}

%\end{titlepage}

\newpage

\section{Introduction and summary}

To quantize an electromagnetic field one could fix a gauge by imposing an   
operator constraint on physical states of the theory. 
For instance, in the Gupta-Bleuler approach  
one postulates
\beq
G(A) |\Psi \rangle =0\,,
\label{GB}
\eeq
where $G(A)$ denotes a function of the gauge field 
operator $A$ or  its derivative, and $ |\Psi \rangle$  
is an arbitrary  {\it physical} state of the theory. 

Alternatively, one could choose to impose the constraint on 
a gauge field already in a 
classical theory in such a way that (\ref {GB}) is 
automatic upon quantization \cite {Nakanishi,Lautrup}. 
This can be achieved, e.g., by introducing  into the classical 
action of the theory a Lagrange multiplier $\lambda $ times 
the function $G(A)$
\beq
\int d^4x \,\lambda(x)\, G(A(x))\,.
\label{lm}
\eeq
The field $\lambda$ has no kinetic or potential terms.
Variation of the action 
w.r.t. $\lambda$ gives a constraint $G(A(x))=0$, 
which is just a classical counterpart of (\ref {GB}).
Small fluctuations of the fields in this theory can consistently be 
quantized for  Abelian \cite{Nakanishi,Lautrup,Symanzik} 
as well as non-Abelian  and gravitational fields 
\cite {Kugo}. The resulting theories can be completed to be 
invariant  under the Becchi-Rouet-Stora-Tyutin (BRTS) transformations 
\cite {BRST}, and a Hilbert space of physical states
can be defined by requiring that the  states carry  
zero BRST and ghost charges \cite {Kugo}.
Quantum effects in the resulting theory are identical to 
those of the Gupta-Bleuler approach.

Nevertheless, there is one difference in using the 
classically constrained theory that has not been explored.  
This difference could be seen in classical 
equations of motion. Variation  of the action  w.r.t. the gauge fields gives  
an equation of motion in which there are new terms proportional 
to $\lambda $ and/or its derivatives. Hence, classical  
field equations are modified, and, depending on 
allowed boundary conditions, new solutions could emerge.

This becomes especially important for  gravity.
General Relativity (GR) with a non-zero cosmological
constant does not admit Minkowski space as a  solution of equations
of motion.  We will discuss  classically constrained General Relativity 
(CGR)  in section 3 and show that the latter does admit flat space
as a solution even if the cosmological constant 
is not zero. This difference is clearly very important.

Unimodular gravity (UGR) \cite {uni} is an interesting 
example of a {\it partially constrained} theory.
In UGR the full reparametrization invariance of GR is 
restricted to a subgroup of volume preserving 
transformations. One practical difference between  
UGR and  GR is that the 
cosmological constant problem in UGR is somewhat relaxed. 
This is because  UGR with a cosmological constant  
admits an infinite number of maximally symmetric solutions
labeled by the value of the space-time curvature. 
However, this  does not explain why one should choose the desirable 
(almost) flat solution among a continuum of maximally 
symmetric ones. Another issue in  UGR 
is related to quantum loops of matter and gravity. 
It is likely that in this theory the Lagrange multiplier $\lambda$ 
acquires quadratic terms via the quantum loops (see, section 3); 
if so, new interactions would be needed to maintain the classical solutions  
of UGR in a quantum theory.

Constrained GR improves on  the above-mentioned 
aspects: (i) it admits only {\it two} maximally 
symmetric solutions -- one with zero curvature, and another one 
with the curvature obtained in GR; (ii) its classical properties, 
due to the  BRST invariance of that model, 
are not modified by the quantum loop effects. 

Reparametrization invariance in CGR is completely constrained.
We will show in section 3, that one can still define a 
{\it locally inertial} reference frame in a small region 
around an arbitrary space-time point. This is because the 
constraint that fixes gauge in the whole space,
allows, in the neighborhood of a given point,
for the point-dependent  gauge transformations 
that locally eliminate  effects of gravity.
Thus the equivalence principle is preserved.

Do  the constrained theories solve the ``old'' cosmological constant problem 
\cite {Weinberg}?  In UGR the answer is negative because
the theory admits an infinite number of maximally symmetric 
solutions. Clearly, in CGR,  there still exists  
a conventional de Sitter solution of GR which can  
be used for inflation in the early universe. 
But the question is whether there exists  an infinite number of 
other  non-maximally  symmetric  solutions in the theory with a 
cosmological constant. If some of  these solutions are 
physical, one should understand why in our Universe the (almost) 
flat solution is preferred. If, on the other hand, the non-maximally symmetric 
solutions can be disregarded for one reason or other, 
then  CGR could be a good starting point for trying to accommodate 
inflation  in the early universe and  still solve  the ``old'' 
cosmological constant problem. These important issues, including the 
question of stability of the new solutions,  are being  studied 
in  \cite {GShang}.  The purpose of the present work is 
to investigate  CGR to determine whether it is a 
consistent low-energy  quantum field theory, which by itself is a
legitimate theoretical question. Concrete applications of this 
model will be discussed in \cite {GShang}.

Even in the most optimistic scenario for CGR, 
one should explain why the space-time curvature 
is not exactly zero but $\sim H_0^2 \sim (10^{-42}~{\rm GeV})^2$, as 
suggested by recent observations.  Where could this  scale 
come from?  One possibility is to introduce a 
pseudo Nambu-Goldstone boson potential 
\cite {Hill} that gives rise to  required ``dark energy''
\footnote{In this case though,  
one needs  a VEV of a scalar to be  somewhat 
higher than the Planck mass.}. Another way  
is to try introducing  a graviton mass $m_g\sim H_0$.  
Although both of the above  
approaches postulate the existence  of a new small scale, 
this scale is stable w.r.t. quantum corrections
(i.e., it is technically natural). In this regard, we briefly 
discuss in the present work 
massive deformations of classically constrained gauge and 
gravitational theories. We will find that the UV behavior of the 
propagators of massive gauge and gravitational quanta are softened.
Although these results are encouraging, at this stage 
we still lack an understanding  of whether the non-linear 
unitarity of the S-matrix on a Hilbert space of physical states 
can be preserved using  the BRST and ghost charges of  these models,
or whether some further modifications may be needed. 
Studies on these issues will be reported elsewhere.

The work is organized as follows: in section 2 we discuss 
a constrained theory of a photon and its  
classical equations  and solutions. 
In section 3 we discuss a classically constrained theory of gravity 
(CGR). We find that CGR with  a cosmological constant has a remarkable
property -- it admits two maximally symmetric 
solutions  one of which is a flat space. We 
find general cosmological solutions in CGR as well as
the expression for the Schwarzschild metric.
An important question that is also addressed in section 3
is that of radiative stability. Using the BRST 
invariant version of the theory we argue  that quantum corrections do not ruin 
the obtained classical results. Massive deformations of CGR are 
also briefly discussed in section 3. In Appendix A 
we discuss the spectrum of the constrained Abelian gauge theory 
in various  approaches, including St\"uckelberg's method. 
We also look at the massive deformation of this model
pointing out that the massive propagator, unlike in the Proca theory, 
has smooth UV behavior and a nonsingular massless limit.
Appendix B  deals with constrained non-Abelian gauge fields.
After briefly discussing classical equations, we 
study the spectrum of the theory. Comments on massive 
deformations of the non-Abelian theories are also included. 
The results are similar to those of the Abelian case.
Some parts of Appendix  A and B are of a review character, but we felt
that including these discussions would make our presentation more 
complete.

\section{Constrained photon}

As an instructive  example we consider  electrodynamics 
with an imposed classical  constraint. We call this model constrained 
QED (CQED) even though we will only quantize it later. We start 
with the Lagrangian density
\beq
\l=-\frac{1}{4}F_{\mu\nu}F^{\mu\nu}
\,+\,A_\mu J^\mu\,+\, \lambda (\partial_\mu A^{\mu} )\,.
\label{qed1}
\eeq
Here $\mu,\nu=0,1,2,3;$ $J^\mu$ is a current, 
and  $\lambda$ is a Lagrange multiplier 
(our choice of the Lorentzian signature is ``mostly negative'').

We start by studying classical properties of this theory. 
The equations of motion that follow from 
the above Lagrangian are
\beq
\partial^\nu F_{\nu\mu}+ J_\mu -\partial_\mu \lambda =0 \,, 
\label{eq1}\\
\partial_\mu A^{\mu}\,=\,0\,.
\label{eq2}
\eeq
Taking a derivative of Eq. (\ref {eq1}) one obtains the following relation
\beq
\square \lambda = \partial^\mu J_\mu \,.
\label{lJ}
\eeq
If the current $J_\mu$ is conserved, $\lambda$ is a harmonic 
function, $\square \lambda =0$. 
One particular solution of this equation is $\partial_\mu \lambda =C_\mu$, 
where $C_\mu$ denotes an arbitrary  space-time constant four-vector.  
Physically, the effective  conserved current to which the gauge 
field is coupled in (\ref {qed1}) is
\beq
 J^{\rm eff}_\mu = J_\mu - \partial_\mu \lambda =J_\mu - C_\mu \,. 
\label{sol}
\eeq
The last term on the r.h.s. of (\ref {sol}) acts as a constant 
background current density determined by a 
vector $C_\mu$.  Since $C_\mu$ is an integration 
constant, there is a continuous set of $C_\mu$'s  that one 
could choose  from.  Setting the value of $C_\mu$ 
is equivalent of  choosing  the corresponding boundary conditions. 
The equations of motion (\ref {eq1}) and 
(\ref {eq2}) were derived by varying the action corresponding 
to  (\ref {qed1}) with the following boundary conditions
\beq
\delta A_\mu |_{\rm boundary} =0, ~~~\delta \lambda|_{\rm boundary}= 
{\rm finite~function}\,.
\label{boundary1}
\eeq
Therefore, all the solutions should obey
(\ref {boundary1}). To demonstrate that such solutions exist let us 
consider a simple example of a spherically symmetric 
localized charge density for which $J_0= \rho\, \theta (r_0-r)$, $J_i=0$. 
Here $\rho$ is a constant charge density, $\theta(r)$ denotes 
the step function, $r$  is the radial coordinate,
and $r_0$ is the  radius of  the charge distribution. We substitute this
source into the RHS of (\ref {eq1}). In addition we chose a solution for 
the Lagrange multiplier to be $\partial_0 \lambda \equiv C_0=\rho$,
and $\partial_j \lambda \equiv C_j=0$.  For this source   
a solution of Eqs. (\ref {eq1}) and (\ref {eq2}), and the 
corresponding electric field read 
\beq
&A_0& = - {\rho r_0^3\over 3 r} - {\rho r^2\over 6}\,~~~{\rm for} ~~r 
\ge r_0\,,~~~A_0 = - {\rho r_0^2\over 2}~~~{\rm for}~~ r\le r_0\,, \nonumber \\
&{\vec E}& = {\rho \over 3 }\,(1 - {r_0^3\over r^3})\,
\theta (r-r_0) {\vec r}\,. 
\label{expot}
\eeq
The above solution satisfies the boundary conditions  
(\ref {boundary1}) at $r=r_0$.
In the conventional electrodynamics the source 
$J_0= \rho \theta (r_0-r)$, $J_i=0$,  yields an electric field  that is 
well-known and differs from (\ref {expot}). The origin of this 
difference is clear --  in  CQED the quantity  $J_\mu$ that 
is specified in the  action {\it is not  the  whole source} 
producing the gauge field! An additional integration constant appears in 
the equations of motion and the total source is 
$J^{\rm eff}_\mu$ (\ref {sol}).  The above theory reduces to conventional 
electrodynamics when we choose $\lambda=0$. This corresponds to what 
we measure in ordinary experiments. 

Similar  solutions with a nonzero value of the Lagrange multiplier  
will play an important  role for gravity  with a nonzero cosmological 
constant. These will be discussed  in section 3 
(non-Abelian gauge fields are discussed in 
Appendix B). 

So far we have not emphasized the fact  that the Lagrangian (\ref {qed1}) 
is not gauge invariant.  In fact, variation of  (\ref {qed1}) 
under the gauge transformation  $\delta A_\mu =\partial_\mu \alpha(x)$ 
vanishes (up to a surface term) for configurations satisfying 
$\square \lambda =0$, therefore (\ref{qed1}) has gauge invariance when 
the $\lambda $ field is {\it on-shell}, even though the gauge field 
could be off-shell. Given that the  $\lambda $
is not propagating, and remains such at the quantum level (see below),
this suggests  two  physical  
degrees of freedom for a photon, while off-shell there are  
four degrees of  freedom.  This will be established more rigorously below.
As it is shown in Appendix A, the two extra off-shell   degree of 
freedom are decoupled from 
conserved sources and have no relevance in the Abelian case.

The quantum theory of CQED could be approached in a number of  
different ways. One could 
restore first a manifest gauge invariance of (\ref {qed1}) 
using the St\"uckelberg method, and then quantize the resulting
gauge invariant theory. We will discuss this in 
Appendix A and B for Abelian and non-Abelian gauge fields respectively.

In the Feynman integral formulation one could think of the 
constrained approach as follows: the gauge and auxiliary fields
can be decomposed in their classical and quantum parts,
$A=A_{cl}+\delta A$, and $\lambda =\lambda_{cl} +\delta \lambda$.
For classical solutions, $A_{cl}$ and $\lambda_{cl}$,
we allow boundary conditions that are different from the conventional ones.
In particular, we allow  for a nonzero solution of the 
$\square \lambda =0$ equation, a zero-mode. 
This is an unconventional step, since  the  usual FP term 
that appears in the path integral is the determinant
of the operator acting on $\lambda$, that is 
$det (\square)$ in this particular case. 
Because of the zero-mode, the determinant, $det (\square)$, 
would have been zero, leading to an ill-defined 
partition function. However, a right way to 
formulate the path integral is to separate
the zero-mode, and  integrate only  w.r.t. the small 
fluctuations for which $det (\square)$  is non-zero, and for which 
the conventional radiation boundary conditions are imposed.

This is the procedure that we will be assuming throughout the text.
Then, quantization of the fluctuations $\delta A$ and $\delta \lambda$ 
over the classical background $A_{cl}$, $\lambda_{cl}$ can be 
performed in a BRST invariant way. That is, one could  start by 
postulating  BRST  invariance of the Lagrangian for the fluctuations  
by adding in the Faddeev-Popov (FP) 
ghosts  without any reference to a local gauge symmetry,
but elevating  BRST invariance to a fundamental  guiding principle in 
constructing the Lagrangian. The resulting BRST symmetric theory of 
small fluctuations could be quantized in a conventional way. 
This is what we summarize below.
For simplicity of presentation we will replace $\delta A_\mu$  
and $\delta \lambda$  by  $A_\mu$ and $\lambda$ keeping in mind that 
these are fluctuations over a classical background (the same replacement 
will be  assumed for non-Abelian and gravity fields considered 
in the next sections and Appendices).

In a conventional Feynman integral approach 
the measure in the path integral 
should be modded out by the gauge equivalent 
classes. The FP trick does this job by introducing
the  gauge-fixing  term along with the FP ghosts. 
In this regard the  following  natural question arises --
since (\ref {qed1}) (or its non-Abelian counterpart) is not 
gauge invariant  why do we need to introduce the FP ghosts in 
the theory?  Naively, it would seem that we should  use  the path 
integral 
\beq
\int DA \, \delta(G(A)) \,{\rm exp }\left ( i \int \l_{\rm Gauge~fields} 
\right ) \,,
\label{path1}
\eeq
instead of the one with the FP ghosts
\beq
\int DA \, \delta(G(A)) \,{\rm det} 
\left | {\delta G^\omega\over \delta \omega }\right |\,
{\rm exp} \left ( { i \int \l_{\rm Gauge~fields} }\right )  \,,
\label{path2}
\eeq
where $G$ is a gauge-fixing condition (for instance it could be that 
$G= \partial_\mu A^\mu $), and $G^\omega$ is 
gauge transformed $G$ with $\omega$ being the transformation 
parameter (the above definition is consistent as far as 
we do not include the zero modes in the integration and 
$det (\square)$ is nonzero, as we discussed above). 
In QED the difference is irrelevant
because the FP ghost are decoupled from the rest of the physics, however,
in the case of non-Abelian fields (\ref {path1}) would lead to a non-unitary 
theory.  Because of the absence of gauge invariance in (\ref {qed1}),
the argument to introduce the FP ghosts in the path integral of  
(\ref {qed1}) cannot be the same as in the conventional case.
Nevertheless, the FP ghosts can be motivated by a symmetry argument, 
namely by requiring the BRST invariance of the constrained 
action.

Let us  modify (\ref {qed1}) by adding the 
FP ghost $c$ and anti-ghost  ${\bar c}$ which are 
Grassmann variables,  $c^2=({\bar c})^2=0$ and ${\bar c}^+ ={\bar c},
~c^+=c$. The Lagrangian reads as follows
\beq
\l=-\frac{1}{4}F_{\mu\nu}F^{\mu\nu}+\, A_\mu J^\mu\, + 
\, \lambda (\partial_\mu A^{\mu} )\,
-\,i {{\bar c}} \square {c}.
\label{qed1ghost}
\eeq
Since on the classical backgrounds considered here the FP ghost fields 
vanish, the classical properties of (\ref {qed1}) and (\ref {qed1ghost})
are identical\footnote{In general, a ghost-antighost bilinear  
could  have nonzero expectation  values on certain 
states, however,  these stats do not  satisfy the zero 
BRST and Ghost charge conditions, see below.}.
The Lagrangian  (\ref {qed1ghost}), however, is invariant 
under the following continuous BRST transformations:
\beq
\delta A_\mu = i \zeta \partial_\mu c\,,~~~
\delta c = 0,~~~\delta {\bar c} =  \lambda {\zeta}\,,~~~\delta\lambda=0\,,
\label{BRST}
\eeq
where $\zeta$ is a  coordinate independent Grassmann 
transformation parameter such that $(\zeta c)^+ = c^+\zeta^+ = c\zeta $. 
The Lagrangian (\ref {qed1ghost}) gives  the 
path integral that is identical to that of the conventional approach
(\ref {path2}) which takes the form
\beq
 \int \fD A \fD \lambda \fD {\bar c} \fD c
\exp\left[i\int -\frac{1}{4}F_{\mu\nu}F^{\mu\nu}+\, A^\mu J_\mu+      
\, \lambda (\partial_\mu A^{\mu} )\,
-\,i {{\bar c}} \square {c}\right ]\,.
\label{pathFP}
\eeq
Notice that the rationale for writing down  (\ref {pathFP}) and 
the Lagrangian (\ref {qed1ghost}) is different from the motivation
that led to the path integral (\ref {path2}).  
In the  constrained approach  the rules 
for constructing a  Lagrangian and path integral are:

(1) Impose classical constraints on fields using  the Lagrange multiplier
technique. Find the corresponding zero-modes and treat them separately
from small fluctuations of the fields. 

(2) Introduce the FP ghosts to obtain the BRST invariance of 
the gauge non-invariant theory. 

(3) Use this Lagrangian to set up the path integral in a 
straightforward way.

The presence of the BRST symmetry guarantees that 
all the Ward-Takahashi identities (the Slavnov-Taylor identities in the
non-Abelian case) of the conventional theory 
are  preserved in the constrained approach,
even though the classical equations of motion
in this approach are different as discussed above.

Let us now turn to the loop corrections that emerge 
in (\ref {qed1ghost}). In particular we would like to
make sure that no kinetic or potential terms are  
generated for $\lambda$. We do have a symmetry $\lambda(x) \to 
\lambda(x) + \beta(x)$, where $\beta$ is an arbitrary function,
w.r.t. which (\ref {qed1ghost}) is invariant. Kinetic or potential 
terms of $\lambda$ would break it.  The question is whether this 
symmetry is preserved by the loop corrections. To address this issue 
we calculate the propagator of the gauge fields.   This can be done 
in a few different ways.  The easiest one is to add a fictitious 
term $-\frac{1}{2}\gamma \lambda^2$  to the Lagrangian 
density and then take the limit $\gamma\rightarrow 0$
\begin{equation}
\begin{split}
Z[J_\mu]\propto &\lim_{\gamma \rightarrow 0}
\int \fD A \fD \lambda \fD {\bar c} \fD c
\exp\left[i \int \l_g+A^\mu J_\mu
+\lambda(\partial_\mu A^\mu)-\frac{1}{2}\gamma \lambda^2 -
i {{\bar c}} \square  {c}  \right  ] \\
\propto &\lim_{\gamma \rightarrow 0}
\int \fD A \fD {\bar c} \fD c
\exp\left[i \int \l_g+A^\mu J_\mu
+\frac{1}{2\gamma }(\partial_\mu A^\mu)^2 -
i {{\bar c}} \square {c}   \right],
\end{split}
\end{equation}
where $\l_g \equiv -\frac{1}{4}F_{\mu\nu} F^{\mu\nu}$.
The propagator is obtained following the standard procedure and 
the result is
\begin{equation}
\label{eq:propagator_CQED}
\Delta^{\mu\nu}= -\lim_{\gamma \rightarrow 0}
\frac{\eta^{\mu\nu}-(1+\gamma )\partial^\mu\partial^\nu/\square}
{\square}=-\frac{\eta^{\mu\nu}-\partial^\mu\partial^\nu/\square}{\square},
\end{equation}
which coincides with the standard transverse 
QED propagator in Landau gauge\footnote{
Here and below we always assume the Feynman (causal) prescription
for the poles in the propagators.}. It is straightforward to check that 
loop corrections preserve the transversality of the 
gauge field propagator.  All the diagrams that renormalize the propagator 
consist of the standard bubble diagrams produced by  the 
contractions between two currents. 
Because the current to which the gauge field is coupled is conserved, the 
propagator remains transverse to all loops
\beq
\partial^\mu \lambda \, \langle A_\mu \,A_\nu \rangle \partial ^\nu \lambda
\sim \partial^\mu \lambda (\eta_{\mu\nu} \square -\partial_\mu \partial_\nu)
\partial ^\nu \lambda =0 \,.
\label{gener}
\eeq 
As a result,  loops  cannot generate the kinetic term for $\lambda$.

Last but not least, the BRST symmetry of (\ref {qed1ghost}) 
can be used to define the Hilbert space of physical 
states  by imposing the standard zero BRST- and Ghost- charge conditions 
$ Q^{\rm BRST}| {\rm Phys}> =0$, $ Q^{\rm Ghost}| {\rm Phys}> =0$. 
In the Abelian case this  reduces \cite {Kugo} to the Nakanishi-Lautrup 
condition $ \lambda^{(-)} | {\rm Phys}> =0$ \cite {Nakanishi,Lautrup}
(here $\lambda^{(-)}$ denotes a negative frequency part of 
a fluctuation of $\lambda$ over a classical background) 
which ensures that  the quantum of $\lambda$ is not in  a set of physical 
{\it in} and {\it out} states of the theory.

\newpage

\section{Constrained models of gravity}

There exists a modification of GR, unimodular gravity (UGR) 
\cite {uni},  that {\it partially}
restricts its gauge freedom and yet reproduces the observable 
results of the Einstein theory. In UGR reparametrization invariance is 
restricted  to the volume-preserving  diffeomorphisms 
that keep the value of $\sqrt{g}$ intact ($g\equiv |det g_{\mu\nu}|$). 
Such a theory can be formulated by using 
the  Lagrange multiplier $\lambda$ (we set $8\pi G_N=1$, unless
indicated otherwise):
\begin{equation}
\mathcal{L}= - { \sqrt{g} \over 2} R - 
\lambda(\sqrt{g}-1)+\mathcal{L}_M,
\label{uni}
\end{equation}
where $\mathcal{L}_M$ is the Lagrangian of the matter fields.
The equations of motion and  
Bianchi identities of this theory
require $\lambda$ to be an arbitrary space-time constant \cite {uni}.  
As a result, UGR is equivalent at the classical level to GR except 
that cosmological term becomes an arbitrary integration  constant. 
Since the latter can take any value, there are an infinite 
number of solutions parametrized by a constant scalar curvature.
Such a theory, at least at the classical level,  
seems to be more favorable than GR -- for  
an arbitrary large value of the vacuum energy in the Lagrangian
(arising, e.g., from particle physics) one is always able  to 
find an almost-flat solution.  Of course this still does not 
explain why one should choose the desirable 
almost-flat solution among a continuum of maximally 
symmetric ones.   This is one aspect of UGR on which one 
would like to improve.

Another, perhaps more pertinent question in UGR emerges when one 
considers quantum loops of matter and gravity. Inspection of 
loop diagrams (see section 4.2) suggest that the  Lagrange 
multiplier $\lambda$ in (\ref {uni}) would  
acquire the mass ($\lambda^2$) as well as 
kinetic ($(\partial_\mu \lambda)^2$) terms due to 
the quantum effects. As a result, $\lambda$
would cease to be an auxiliary field, and all the classical 
results of UGR would have to be reconsidered.

We will discuss a model that completely constrains  
reparametrization invariance of GR.  This theory has the following 
two important properties: (I) It allows, like UGR does,  
an adjustment of the cosmological constant via the integration 
constant mechanism, but only admits two maximally symmetric
solutions, one of which is a flat space.
(II) Unlike UGR its classical properties are stable w.r.t. 
quantum corrections, i.e., the Lagrange
multiplier remains an auxiliary field even in the 
quantum theory. This theory also preserves the equivalence principle.

\subsection{Constrained gravity and the cosmological constant}

Consider the following Lagrangian 
\begin{equation}
\mathcal{L}=-{\sqrt{g}\over 2} R
+\sqrt{g} g^{\mu\nu}\partial_{\mu}\lambda_\nu+\mathcal{L}_M +
{\rm surface~terms}.
\label{congrl}
\end{equation}
Here, $\lambda_\nu$ is a vector that serves as a Lagrange multiplier and 
$\mathcal{L}_M $ denotes the Lagrangian of other fields which
can also include  a  vacuum energy term (the cosmological constant)
produced by classical and/or quantum effects. Versions of 
this model have been discussed in the literature previously  
(see, e.g., the last reference in \cite {Kugo}) 
with the purpose of  introducing the de Donder  
gauge-fixing condition in the context of 
quantization of gravity. Here, instead,  we regard this model  
as a classical theory, which is subsequently quantized, but  
the classical equations of which could admit rather interesting 
solutions that are absent in GR. Thus, 
we first concentrate on classical effects,
leaving the discussion of quantum corrections 
for the next subsection\footnote{One could choose to 
impose a different  constraint using the Lagrange multiplier 
in (\ref {congrl}) (for instance, an axial-gauge constraint).
As long as it is an acceptable gauge-fixing condition
for small fluctuations, and the corresponding FP ghosts are taken 
care of consistently  (in the axial-gauge the FP ghost are not needed) 
the quantum effects of the fluctuations won't depend on the choice of 
this constraint.  However, the classical solutions could differ for 
different constraints. Here we choose the constraint that is 
Lorentz-invariant.}.

Among others, we will be considering below 
solutions with fixed boundaries. For such solutions 
one should add to the  Einstein-Hilbert action the Gibbons-Hawking  
boundary term. Moreover, the boundary conditions that we allow for are:
\beq
\delta g_{\mu\nu} |_{\rm boundary} =0, ~~{\rm and}~~
\delta \lambda_\mu |_{\rm boundary}= 0\,.
\label{boundaryg1} 
\eeq
Under these conditions the variation w.r.t. the metric gives 
\beq
G_{\mu\nu}- \left(\partial_\mu\lambda_\nu
+\partial_\nu\lambda_\mu\right)+ g_{\mu\nu} \partial^\sigma 
\lambda_\sigma =T_{\mu\nu}\,. 
\label{eqcon1} 
\eeq
In addition to this, we get a constraining equation
by varying the action w.r.t. the Lagrange multiplier
\beq
\partial_{\mu}(\sqrt{g}\,g^{\mu\nu})=0\,.
\label{har}
\eeq
The above model is similar in spirit to a constrained 
gauge theory  discussed  in the previous section. 
It is easily checked that  (\ref {har})  is equivalent to 
the condition 
\beq
\sqrt{g}\Gamma^\alpha_{\mu\nu}g^{\mu\nu}=0\,,
\label{har1}
\eeq
and, in the linearized
approximation, gives rise  to 
de Donder (harmonic) gauge fixing of linearized 
GR. Due to this condition, $\nabla_\mu A^\mu =g^{\mu\nu}\partial_\mu A_\nu
\equiv \partial^\mu A_\mu$, where $\nabla_\mu$ is a 
covariant derivative acting on  an arbitrary four-vector $A_\mu$. 

The constraint  (\ref {har}) (or (\ref {har1}))  
fixes completely reparametrization 
invariance of the theory. How is then the equivalence 
principle recovered? 
A relevant property of (\ref {har}) (or (\ref {har1})) is 
this: for any given point $x^\mu =x^\mu_0$ in the coordinate system 
$\{x^\mu\}$,  it allows for the $x^\mu_0$-dependent coordinate 
transformations that eliminate the connection  in a small neighborhood 
of this point. These transformations can be written as 
\beq
x^{\prime \mu} = x^\mu + {1\over 2}\Gamma^\mu_{\alpha\beta}(x_0)
(x-x_0)^\alpha (x-x_0)^\beta\,.
\label{trans}
\eeq
It is straightforward to check that for (\ref {trans})
$g^{\alpha\beta}(x)|_{x=x_0}=
g^{\prime \alpha\beta}(x^\prime)|_{x^\prime=x_0}$, and 
\beq
\Gamma^\mu_{\alpha\beta}(x)|_{x=x_0} = \Gamma^{\prime \mu}_
{\alpha\beta}(x^\prime)|_{x^\prime=x_0}+ 
\Gamma^\mu_{\alpha\beta}(x)|_{x=x_0}\,.
\label{ch}
\eeq 
As a result, $ \Gamma^{\prime \mu}_
{\alpha\beta}(x^\prime)|_{x^\prime=x_0}=0$, and at this point 
the metric can simultaneously be brought to the Minkowski form.  
The transformation  (\ref {trans}), 
on the surface of (\ref {har}) (or (\ref {har1})), 
is trivially consistent with (\ref {har}) (or (\ref {har1}))  
\beq
g^{\alpha\beta}\Gamma^\mu_{\alpha\beta}(x)|_{x=x_0} = 
g^{\prime \alpha\beta}(x^\prime)
\Gamma^{\prime \mu}_{\alpha\beta}(x^\prime)|_{x^\prime=x_0}\,=0.
\label{cons125}
\eeq
Since $x_0$ was  arbitrary, 
the above arguments can be repeated for  any other point in 
space-time. One should emphasize  again 
that the coordinate transformation (\ref {trans}) 
is {\it point-specific}, i.e., at different points 
of space-time one should perform transformations that 
depend  parametrically on that
very point. This is why they are not gauge transformations in 
the entire space-time. Summarizing, although 
(\ref {har}) (or (\ref {har1})) picks a global  coordinate
frame, it allows for the point-dependent coordinate 
transformations that can eliminate gravity locally. 
Hence, the equivalence principle. The local Lorentz 
transformations are also preserved.

To obtain  the equation which the  Lagrange multiplier 
has to satisfy we  apply  a covariant derivative to both sides  
of (\ref {eqcon1}). This is subtle since the second term on the l.h.s.
of (\ref {eqcon1}) is not a tensor and we should 
define the action of a covariant derivative on this object.
We adopt the following straightforward procedure: 
apply to both sides of (\ref {eqcon1}) the operator
\beq
\nabla^\alpha \equiv g^{\alpha \mu} \left 
( \partial_\mu  - \Gamma^{*}_{*\mu} -\Gamma^{*}_{*\mu} \right ),  
\label{op}
\eeq
where the standard index arrangement
should be used in place of the  
asterisks even if the two-index object on which this operator is acting
does not transform as a tensor.  
Then, using the Bianchi identities and covariant 
conservation of the stress-tensor we obtain:
\begin{equation}
g^{\mu\alpha}\nabla_\mu 
(\partial_\alpha\lambda_\nu+\partial_\nu\lambda_\alpha)
=\nabla_\nu \partial^\sigma \lambda_\sigma\,.
\label{lambdaeq}
\end{equation}
The above equation can be simplified substantially
due to (\ref {har}). The left hand side of (\ref {lambdaeq}) 
can be reduced to 
\beq 
g^{\mu\alpha} 
\partial_\mu \left(\partial_\alpha\lambda_\nu+
\partial_\nu\lambda_\alpha\right) -
g^{\mu\alpha}\left(\partial_\alpha\lambda_\beta+
\partial_\beta\lambda_\alpha\right) 
\Gamma^{\beta}_{\nu\mu}\,,
\nonumber
\eeq
while the right hand side simplifies to give
\beq 
(\partial_\nu g^{\alpha\beta})\partial_\alpha\lambda_\beta
+g^{\alpha\beta}\partial_\nu\partial_\alpha\lambda_\beta .
\nonumber 
\eeq
Combining the above two expressions together we find from
(\ref {lambdaeq}) 
\begin{equation}
g^{\mu\alpha}\partial_\mu\partial_\alpha \lambda_\nu=0 \,.
\label{simple}
\end{equation}
As long as $g^{\mu\nu}$ is non-singular general solutions
for $\lambda_{\nu}$ could be found. 

Clearly, the system of equations (\ref {eqcon1}), (\ref {har})
and (\ref {lambdaeq}) (or (\ref {simple})), 
could admit new solutions that are absent in the Einstein theory.
For instance, the Einstein equations with a nonzero 
cosmological constant (i.e., $T_{\mu \nu}= \Lambda g_{\mu\nu}$), 
do not admit Minkowski space as a solution. In contrast with this, 
the system (\ref {eqcon1}), (\ref {har}) and (\ref {lambdaeq})
is satisfied by the flat space metric
\beq
g_{\mu\nu} =\eta_{\mu\nu},~~~
\partial_\mu \lambda_\nu + \partial_\nu \lambda_\mu =  \Lambda \,
\eta_{\mu\nu}\,.
\label{Mink}
\eeq
It is remarkable that one can obtain a flat solution even though the 
vacuum energy in the Lagrangian is not zero!  Similar property exists 
in unimodular gravity \cite {uni} when $T_{\mu \nu}= \Lambda g_{\mu\nu}$,
but there are an infinite number of maximally symmetric solutions labeled 
by the value of a constant curvature.  Is  this also true in the model 
(\ref {congrl})?  We analyze this issue below.
First we notice that the system (\ref {eqcon1}), (\ref {har})
and (\ref {lambdaeq}) admits a de Sitter solution 
of conventional general relativity (we will focus on the case
of a positive cosmological constant only from now on and interesting 
results can  be obtained for both positive and negative 
$\Lambda$, \cite {GShang}).  The conventional dS metric 
solves (\ref {eqcon1}) with $\lambda_\mu =const.$. What is less obvious 
is how this solution satisfies (\ref {har}). To understand this 
we start with the  dS solution in the co-moving coordinate system
(this is also applicable to any FRW cosmology) 
\begin{equation}
\ud s^2=\ud t^2-a^2(t)\left(\ud x^2+\ud y^2+\ud z^2\right).
\label{dSco}
\end{equation}
As it can be checked directly, (\ref {dSco}) does 
not satisfy (\ref {har}). However, we can define a new 
time variable  
\begin{equation}
\tau\equiv \int \frac{\ud t}{a^3(t)}\,,
\label{eta}
\end{equation}
for which the interval becomes
\begin{equation}
\ud s^2=a^6(\tau) \ud \tau^2-a^2(\tau)(\ud x^2+\ud y^2+\ud z^2).
\label{dSint}
\end{equation}
This metric satisfies (\ref {har}) identically.
Therefore, a conventional dS metric,
or any other spatially-flat FRW cosmology, is a solution of the system 
(\ref {eqcon1}), (\ref {har}), (\ref {lambdaeq}).

Having established that the flat and conventional dS 
spaces are two solutions of the theory,
let us now look at other possible maximally symmetric solutions.
In general, the following ansatz
\beq
\partial_\mu \lambda_\nu + \partial_\nu \lambda_\mu = c g^{dS}_{\mu\nu}\,, 
\label{maxsym}
\eeq
where $c$ is an arbitrary constant, $g^{dS}_{\mu\nu}$ is a dS 
solution with $ R=-4(\Lambda-c)$,  does satisfy the equations 
(\ref {eqcon1}), (\ref {har}) and (\ref {lambdaeq}).
However, this is not  enough to claim that (\ref {maxsym})
is a legitimate solution of the theory.  This is because 
equation (\ref {maxsym})  itself might not be solvable in terms 
of $\lambda_\mu$ given that  $g_{\mu\nu}$ is a dS metric obeying
(\ref {har}); solvability for $\lambda_\mu$  
is a necessary condition since it is w.r.t. $\lambda_\mu$ 
that we varied the action.  It is straightforward to check that 
there is no solution for $\lambda_\mu$ that would satisfy equation 
(\ref {maxsym}) if the metric is given by (\ref {dSint}).
Could there be other forms of dS space that are 
non-trivially different from (\ref {dSint}) and yet satisfy
(\ref {har})? The answer is no. To see this 
consider a dS solution in GR in two different coordinate 
systems. Let us {\it assume} the opposite, 
that both of these coordinate systems can be 
gauge transformed in GR to two different  
coordinate systems for which (\ref {har}) is valid. 
If true, this would mean 
that the condition (\ref {har}) does not completely
fix the gauge freedom of GR. On the other hand, we find
by performing gauge transformation of (\ref {har}) that this is 
only possible if the gauge transformation itself is trivial. 
Therefore,  the form (\ref {dSint})
is the unique dS solution which satisfies the constraints (\ref {har}).

The fact that (\ref {maxsym}) cannot be a maximally symmetric 
solution with nonzero $R(g)$ could also be established just 
by looking at a general expression of the Ricci scalar in terms of 
the metric $g_{\mu\nu}$ in which the substitution 
$c g_{\mu\nu}= \partial_\mu \lambda_\nu + \partial_\nu \lambda_\mu$
is made.  If this ansatz could describe a dS space, one could always 
go to a weak-field regime where  the Ricci curvature, as a function of 
the metric,  has to be nonzero. However, the above ansatz for $g_{\mu\nu}$ 
gives zero $R$ in the leading order of the weak-field approximation.

Summarizing, we conclude that in the class of maximally symmetric 
spaces there are only two solutions of the theory: 
(I) The flat space defined in (\ref {Mink}); 
(II) The  (A)dS solution as it would appear in 
conventional GR transformed to a new coordinate system.
There could in principle exist other, non-maximally symmetric 
solutions. It would be interesting to study whether those solutions
are physical. If they are one should 
look for arguments why the maximally symmetric solutions could be  
preferred in our Universe. On the other hand, if the non-maximally-symmetric  
solutions  are not there, then the model (\ref {congrl}) 
could  be a playground for studying the fate of the  
``old'' cosmological constant problem  \cite {Weinberg}. 
It is interesting to point out that  by integrating out 
the Lagrange multiplier field one gets a non-local action, with some 
similarities but also differences of the resulting non-local theory 
with that of Ref. \cite {ADDG0}.

\vspace{0.1in}

We would also point out that the Schwarzschild metric
of conventional GR is also a legitimate solution of 
the CGR. A simplest way to address this is to choose 
$\lambda_\nu =0$ and make sure that the known GR solution 
itself  satisfies (\ref {har}) in a particular coordinate system.
Usual solutions of GR can be transformed to satisfy  Eq. (\ref {har}).
This is in particular true  when $g_{\mu\nu}$
is diagonal and each element of $\sqrt{g} g_{\mu\mu}$ 
(there is no summation w.r.t. $\mu$ here)  factorizes into the 
products of the form $\sqrt{g} 
g_{\mu\mu}=h(x^\alpha)j(x^\nu)\cdots f(x^\lambda)$,
where each function depends on one coordinate only.
In this case, the constraints \eqref{har} turn into four separate 
partial differential equations
\begin{equation}
\partial_\mu (\sqrt{g} g_{\mu\mu}^{-1})=0, \qquad 
\mu=0, 1, 2, 3,\quad \textrm{no summation w.r.t. } \mu.
\label{eq:constraints_diagonal}
\end{equation}
Suppose we introduce one
new coordinate $\tilde x^{\alpha}=\tilde x^{\alpha}(x^{\alpha})$
such that it depends only on $x^\alpha$, and  leave all
the other coordinates intact.
In the new coordinate system
\begin{equation}
\sqrt{-\tilde g}= x'^\alpha \sqrt{-g}\,,
\end{equation}
while 
\begin{equation}
\tilde g^{\alpha\alpha}=(x'^{\alpha})^{-2}g^{\alpha\alpha}\,,~~ \textrm{and}
\quad
\tilde g^{\mu\mu}=g^{\mu\mu}\;~ \textrm{for}\;~ \mu\neq\alpha,
\end{equation}
where we have defined $x'^\alpha\equiv \frac{\ud x^\alpha}{\ud\tilde
x^\alpha}$ and chosen it to be positive.

The $\alpha$-th equation of \eqref{eq:constraints_diagonal}
in the new coordinate system takes the form:
\begin{equation}
\frac{\partial (\sqrt{-\tilde g} \tilde g^{\alpha\alpha})}
{\partial \tilde x^\alpha}
=\frac{\partial [\sqrt{-g} g^{\alpha\alpha} (x'^{\alpha})^{-1}]}
{\partial \tilde x^\alpha}=0\,.
\end{equation}
If  $\sqrt{-g}g^{\alpha\alpha}$ can be factorized, say as,
$\sqrt{-g}g^{\alpha\alpha}=h(x^\alpha)\psi$ where $\psi$ depends only on
coordinates other than $x^\alpha$, we can find the desired
$\tilde x^\alpha$ by simply demanding
\begin{equation}
\frac{h(x^\alpha)}{x'^\alpha(\tilde x^\alpha)}=1\,, \quad \textrm{or any
constant if more convenient,}
\end{equation}
and solving this ordinary differential equation. It is not 
difficult to see that one can carry on the
same procedure for each $x^\mu$'s  without invalidating the previous
results, and, therefore, eventually find the new coordinate system
that satisfies (\ref {har}).

The above procedure is directly  applicable to the Schwarzschild 
and FRW  solutions. For the latter the result was 
already given above (see (\ref {dSint})). Here we perform the change 
of coordinates for the Schwarzschild metric. In a  spherically 
symmetric coordinates 
\begin{equation}
\ud s^2=\left(1-\frac{r_g}{r}\right)\ud t^2
-\left(1-\frac{r_g}{r}\right)^{-1}\ud r^2
-r^2\left(\ud \theta^2 +\sin^2\theta \ud \varphi^2\right)\,.
\end{equation}
Here, $r_g =2 G_N M$ is the horizon radius of an object.
The above described procedure leads to the new coordinate system
\begin{equation}
\begin{split}
\tilde r=r_g\ln{\frac{r-r_g}{r}}~~ \Rightarrow~~& 
r=\frac{r_g}{1-e^{\tilde r/r_g}}\,,\label {changer}\\
\tilde\theta=\ln{\tan \frac{\theta}{2}}~~\Rightarrow~~&
\theta=2\tan^{-1}e^{\tilde\theta}\,,
\end{split}
\end{equation}
in which the Schwarzschild metric becomes
\begin{equation}
\ud s^2=e^{\tilde r/r_g}\ud t^2
-\frac{e^{\tilde r/r_g}}{(e^{\tilde r/r_g}-1)^4}\ud \tilde r^2
-\frac{r^2_g}{(e^{\tilde r/r_g}-1)^2\cosh^2 \tilde \theta}
\left(\ud \tilde \theta^2+\ud \varphi^2\right)\,.
\end{equation}
This metric satisfies (\ref {har}).
The new variable ${\tilde r}$ spans the interval
$(-\infty,~0]$ as the coordinate $r$ increases from $r_g$ 
to $+\infty$, while the new angular variable ${\tilde \theta}$
covers the interval $(-\infty,~+\infty)$. One can also easily describe 
the interior of the Schwarzschild solution by flipping the sign of 
the argument of the log in (\ref {changer}).

\subsection{Radiative stability}

In this section we discuss the issue of 
quantum loop corrections to (\ref {congrl}).  
In particular, we would like to understand  
whether this theory is stable w.r.t. the loops.
It is clear that quantum-gravitational and matter 
loops will generate higher dimensional operators 
entering the action suppressed by the UV cutoff of this theory. 
This is similar to any theory that is not renormalizable and should be 
regarded as an effective field theory below its UV cutoff
(for an introduction to an effective field theory treatment 
of gravity, see, e.g., \cite {Donoghue}).
However, there  is another question of a  vital importance in the 
present context. This is whether loop corrections can generate 
the potential and/or kinetic terms for the Lagrange 
multiplier. If this happens, $\lambda_\mu$  
cannot be regarded as an auxiliary field and all the results
of the previous subsection would be ruined.

We will argue below that this problem is avoided in CGR because of the 
specific form of (\ref {congrl}) which can be completed to a BRST 
invariant theory.  Following Ref. \cite {Gold}, we  introduce 
the variables (as in the previous sections, below we
are discussing small fluctuations of the fields)
\beq
\gamma^{\mu\nu} = \sqrt {g} g^{\mu\nu}\,,~~~
\gamma_{\mu\nu} = { g_{\mu\nu} \over \sqrt {g} }\,.  
\label{gamma}
\eeq
It is straightforward to rewrite  the Lagrangian (\ref {congrl}) 
in terms of these variables and include the FP ghost term:
\beq
{\cal L}= -{1\over 2} \gamma^{\mu\nu}\left ( R_{\mu\nu}(\gamma) -
\partial_\mu \lambda_\nu - \partial_\nu \lambda_\mu \right ) +
{i\over 2} \left (\partial_\mu {\bar c}_\nu +
\partial_\nu {\bar c}_\mu \right ) \nabla^{\mu\nu}_\alpha c^\alpha\,. 
\label{congrl1}
\eeq
Here the terms in the first parenthesis represent  the gravitational
part of (\ref {congrl}), while  the last term introduces
the vector-like FP ghost and anti-ghost fields for which  
the operator $\nabla^{\mu\nu}_\alpha $ is defined as follows:
\beq
\nabla^{\mu\nu}_\alpha \equiv \gamma^{\mu\tau} \delta^\nu_\alpha 
\partial_\tau + \gamma^{\nu\tau} \delta^\mu_\alpha \partial_\tau -
\partial_\alpha (  \gamma^{\mu\nu} ~~~)\,.
\nonumber
\eeq
The important point is that the standard Einstein-Hilbert action
can be rewritten in terms of (\ref {gamma}) and their first 
derivatives \cite {Gold}
\beq
-{1\over 2} \gamma^{\mu\nu} R_{\mu\nu}(\gamma) =
-{1\over 8} \partial_\rho \gamma^{\mu\tau} \partial_\sigma \gamma^{\lambda\nu}
\left (\gamma^{\rho\sigma} \gamma_{\lambda \mu} \gamma_{\tau \nu}
-{1\over 2}\gamma^{\rho\sigma} \gamma_{\mu \tau } \gamma_{\lambda \nu} -
2 \delta^\sigma_\tau \delta ^\rho_\lambda \gamma_{\mu\nu}\right )\,.
\label{ehgamma}
\eeq
The  FP ghost term in (\ref {congrl1}) 
ensures the BRST invariance of this theory \cite {BRSTgrav}.
The respective BRST transformations with a continuous Grassmann 
variable $\zeta$ are:
\beq
\delta \gamma^{\mu\nu} = i\nabla^{\mu\nu}_\alpha c^\alpha \zeta , \nonumber \\
\delta c^\mu = i c^\tau \partial_\tau c^\mu \zeta , \nonumber \\
\delta {\bar c}^\mu = - \lambda_\mu \zeta,~~~~\delta {\lambda}_\mu=0\,.
\nonumber 
\label{BRSTgr}
\eeq
Presence of this symmetry ensures that the 
Lagrange multiplier in (\ref {congrl1}) 
does not acquire the kinetic term through the loop 
corrections. This is because the potentially dangerous 
term in the effective Lagrangian 
\beq
\partial_\mu \lambda_\nu\langle \gamma^{\mu\nu}(x)  \gamma^{\alpha\beta}(0)
\rangle \partial_\alpha \lambda_\beta \,,
\label{renlambda}
\eeq
is zero (up to a total derivatives) due to  the transversality of 
a two point graviton correlation function. The latter can be seen 
order by 
order in perturbation theory.  Let us for simplicity consider this 
for an expansion
about a  flat space which is a consistent solution of the theory 
even if the cosmological constant is present. We introduce the 
notations $\gamma^{\mu\nu} = \eta^{\mu\nu} - \varphi^{\mu\nu}$,
and look at the two-point correlation function of the $\varphi^{\mu\nu}$
field. This correlator has been studied in detail in  the conventional 
approach, in which there is no Lagrange multiplier term 
in (\ref {congrl1}), but instead, a standard
quadratic gauge fixing term ${(\partial_\mu \gamma^{\mu\nu})^2
\over 2\zeta}$ with the gauge parameter $\zeta$ is introduced. 
Our theory, on a fixed classical 
background, corresponds to the limit  $\zeta \to 0$. Hence, 
all the results concerning the quantum loops derived in the 
conventional approach on a fixed background 
are also applicable here with the condition that  
$\zeta \to 0$.  The BRST invariance of the theory 
can be used to deduce the Slavnov-Taylor 
identities \cite {Slavnov} (see, e.g., \cite {Capper,Niu,Kugo}).
The latter 
guarantee  that order by order in perturbation theory the 
two point correlation function of the  $\varphi^{\mu\nu}$
field is transverse  (this corresponds to the 
$\zeta\to 0$ gauge results of Refs. \cite {Capper,Niu,Kugo})
\beq
\langle \varphi^{\mu\nu}(x)  \varphi^{\alpha \beta}(0) \rangle
\propto \Pi^{\mu\alpha} \Pi^{\nu\beta} + 
\Pi^{\mu\beta} \Pi^{\nu\alpha} - \Pi^{\mu\nu} \Pi^{\alpha \beta}, \nonumber \\
{\rm where}~~~~~~\Pi^{\mu\nu} \equiv  \eta^{\mu\nu} - 
{ \partial^\mu \partial^\nu\over 
\square} \,. 
\nonumber
\eeq
In conclusion, the presence or absence 
of the FP ghosts does not affect the classical 
equations of motion, and for the classical analysis it is 
acceptable to ignore them and study the Lagrangian (\ref {congrl}). 
The only quantum mechanically consistent theory is that  
with the FP ghosts described by (\ref {congrl1}) and 
all the classical results  obtained above hold in 
this theory.  Furthermore those results are 
stable w.r.t. the quantum loop corrections.

We would like to comment on a similar issue  in the context of 
UGR \cite {uni}. The Lagrangian of this theory, as given 
in (\ref {uni}), is likely to generate quadratic terms 
for $\lambda$ via loops. To see this we start with   
a one-loop diagram of Fig. 1.
\begin{figure}
\centerline{\epsfbox{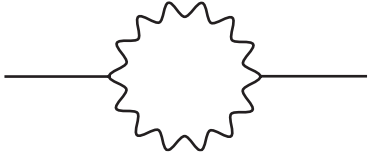}}
\epsfysize=6cm
\vspace{0.1in}
\caption{\small The solid lines denote the $\lambda$ field 
(there is no propagator corresponding to these lines, they 
are depicted to show the vertices);
the wave lines correspond to  gravitons. The vertices  in 
this one-loop diagram arise 
due to the cubic interaction of $\lambda$  with two gravitons 
originating from the term $\lambda \sqrt{g}$ in the Lagrangian (\ref {uni}).} 
\label{1a}
\end{figure} 
This diagram is logarithmically divergent and 
will generate an additional term proportional to 
\beq
{\lambda^2 \over \mpl^4} {\rm log} \left ( {\mu_{UV}\over \mu_{IR}} 
\right )\,,
\label{massL}
\eeq
where $\mu_{UV}$ and $\mu_{IR}$ denote the UV and IR scales respectively,
and we have restored  $\mpl^2$ in front of 
the $\sqrt{g}R$ which resulted in the $1/\mpl^4$ coefficient in 
(\ref {massL}).  Naively, the above term  may seem to be irrelevant
because of the  $1/\mpl^4$ suppression. However, to get a right 
scaling we should restore the canonical dimensionality of $\lambda$. 
This is achieved by substitution $\lambda\to \mpl^3 \alpha$, 
after which the $\alpha$-dependent terms in the effective 
Lagrangian take the form
\beq
\mpl^3 \alpha (\sqrt {g}-1) +\mpl^2 \alpha^2 {\rm log} 
\left ( {\mu_{UV}\over \mu_{IR}}\right ).
\label{uv}
\eeq
Because of the new induced term, the field $\alpha $ acquires a Planckian 
mass  and ceases to be  an auxiliary field. 
Furthermore, we could also look at a two-loop diagram of 
Fig 2.  A simple power-counting of the momenta running in 
the loops shows that this diagram generates not only a mass term for 
$\alpha$ but also its  kinetic term:
\beq
{(\partial_\mu  \lambda)^2 \over \mpl^6} {\rm log} 
\left ( {\mu_{UV}\over \mu_{IR}}\right )\,\to 
{(\partial_\mu  \alpha)^2} {\rm log} 
\left ( {\mu_{UV}\over \mu_{IR}}\right ).
\nonumber
\eeq
Higher loops are also expected to generate similar terms,
and the  above arguments are hard to avoid unless the theory
(\ref {uni}) is amended by new interactions. Perhaps the 
BRST invariant completion of UGR proposed in Ref. \cite {Buch2}
can cure this;  it would be interesting to perform 
explicit calculations in the framework of  \cite {Buch2} 
to see whether the radiative stability is restored.
Since the model of \cite {Buch2}  is BRST invariant one 
would expect a positive outcome. However,  
we should point out that the bosonic part of \cite {Buch2} 
contains additional gauge fixing terms needed to completely restrict 
parametrization invariance of the theory, and, 
from this perspective, it differs from UGR.
\begin{figure}
\centerline{\epsfbox{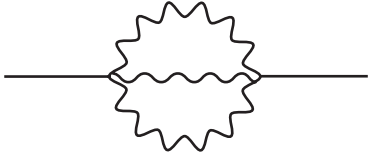}}
\epsfxsize=6cm
\vspace{0.1in}
\caption{\small The solid lines denote the $\lambda$ field 
(there is no propagator corresponding to these lines, they 
are depicted to show the vertices); the wave lines correspond to  gravitons.
The vertices  in this one-loop diagram arise 
due to the quartic interaction of $\lambda$  with three 
gravitons originating from the term $\lambda \sqrt{g}$ in 
the Lagrangian (\ref {uni}).} 
\label{1b}
\end{figure}

Finally, we would like to calculate a response of the 
graviton field to a source. In the linearized approximation 
the number of physical propagating degrees of freedom of CGR 
should be the same as in GR. In the linearized approximation
$\partial_\mu \varphi^{\mu\nu} =0$ due to (\ref {har}),
and $\partial^2 \lambda_\nu =0$ due to (\ref {lambdaeq}).
Then, the equation (\ref {eqcon1}) simplifies to
\beq
\square \varphi_{\mu\nu} = T_{\mu\nu}\,. 
\label{resp}
\eeq 
The above equation is identical to an expression for the response 
in the Einstein theory. This of course is a consequence 
of the fact that the free propagator of (\ref {congrl1}), 
coincides with the graviton propagator of GR in the harmonic gauge 
$ \partial_\mu \varphi^{\mu\nu}=  -\partial_\mu h^{\mu\nu} +{
1\over 2} \partial^\nu h =0$, where
$g^{\mu\nu} \simeq \eta^{\mu\nu} -h^{\mu\nu} $.

\subsection{Comments on massive theories}

It is difficult to construct a consistent Lorentz-invariant 
nonlinear model of a massive 
graviton propagating on a Minkowski background. In the linearized 
approximation the only consistent massive deformation 
of GR  is the  Fierz-Pauli (F-P) model \cite {FP}
\beq
\mathcal{L}=-{\sqrt{g}\over 2} R +\frac{\sqrt{g}}{4}m_g^2
(h_{\mu\nu}h^{\mu\nu}-h^2)\,,
\label{FP}
\eeq
were $h_{\mu\nu}\equiv g_{\mu\nu} -\eta_{\mu\nu} $.
In the quadratic approximation this model describes a 
massive spin-2 state  with 5 degrees of freedom.
However, a  nonlinear completion of this theory is not unique, and
so far there is no  known non-linear theory in four-dimensions 
that would be consistent. 
A rather general class of nonlinear completions of the F-P theory
give rise to unbounded from below 
Hamiltonian \cite {BD}. This manifests itself in classical 
instabilities for  which the time scale can be substantially 
shorter than the scale of the inverse graviton mass
\cite {GGruzinov,Nicolis,Cedric}.

The reason for this instability is that at the nonlinear level
a ghost-like sixth ``degree of freedom'' shows up. This should have 
been expected because of the following.
Ten degrees of freedom of $g_{\mu\nu}$ in (\ref {FP}) are restricted 
only by four independent  Bianchi identities. 
Hence, six degrees of freedom should remain.
The absence of the sixth degree of freedom in the linearized 
theory was just an artifact  of  the linearized approximation 
itself \cite {BD}.

It is interesting to ask the following question:
what if we start with a mass deformation of a constrained
gravitational theory instead of modifying GR as in (\ref {FP})?
Straightforward calculations show  that the mass deformation of 
unimodular gravity (\ref {uni}) or the constrained gravity (\ref {congrl})
leads to a theory with a ghost in the linearized approximation.
This ghost can be removed, at least in the linearized theory, 
if one considers a mass deformation of a model with both  constraints
$\sqrt {g}=1 $ and $ \partial_\mu \sqrt {g} g^{\mu\nu}=0$.
Because this set of equations  imposes 5 conditions 
on ten components of $g_{\mu\nu}$, one should expect this 
theory to propagate  5 physical degrees of freedom. 
Below we will discuss the advantages as well as difficulties 
of this approach.

\vspace{0.1in}

The Lagrangian of the massive ``hybrid model'' that combines the two 
constraints mentioned above takes the form:  
\begin{equation}
%\label{eq:lagrangian_hy}
\mathcal{L}=
-{\sqrt{g}\over 2}R+\frac{\sqrt{g}}{4} m_g^2
(h_{\mu\nu}h^{\mu\nu}-h^2) +\sqrt{g}g^{\mu\nu}\partial_\mu \lambda_\nu
-\lambda(\sqrt{g} -1)\,.
\label{hybrid}
\end{equation}
Variation of the action w.r.t.the Lagrange multipliers  
$\lambda$ and $\lambda_\mu$  yields the constraints:
\begin{equation}
\sqrt{g}=1;~~~\qquad \partial_\mu \sqrt{g}g^{\mu\nu}=0\,.
\label{Hcon}
\end{equation}

We now turn to the linearized approximation about a flat space 
to study  a graviton propagator. In this approximation 
the constraints (\ref {Hcon}) reduce to $h=0$, and 
$\partial^\mu h_{\mu\nu}=0$ respectively. The 
equation of motion becomes
\begin{equation}
\label{eqh}
G_{\mu\nu}-m_g^2 h_{\mu\nu} - \partial_{(\mu}\lambda_{\nu)}
+(\partial^\alpha \lambda_\alpha)\eta_{\mu\nu} 
- \lambda\eta_{\mu\nu} = T_{\mu\nu}.
\end{equation}
The trace equation and the Bianchi  identities give respectively  
\begin{equation}
-4 \lambda +2 \partial^\mu \lambda_\mu = T \,,~~~
\partial_\mu \lambda +\partial^2 \lambda_\mu =0\,,
\nonumber
\end{equation}
which can be solved to obtain
\begin{equation}
\label{eq:lambdas_hy}
%\begin{split}
\lambda_\mu =\frac{\partial_\mu T}{6 \square}\,,~~~~~
\lambda =-\frac{T}{6} \,.
%\end{split}
\end{equation}
Substituting these solutions into (\ref{eqh}) we  find
\begin{equation}
\label{hybh}
\begin{split}
h_{\mu\nu}=  \frac{T_{\mu\nu}-\frac{1}{3}T \eta_{\mu\nu}}{\square +m_g^2}
+\frac{\partial_\mu\partial_\nu T}{3 \square(\square +m_g^2)}\,.
\end{split}
\end{equation}
From the tensorial structure of (\ref {hybh}) we conclude that 
the theory propagates five physical polarizations, as it should.
Moreover, unlike F-P gravity, the propagator (\ref {hybh}) 
has a well-defined $m_g\to 0$ limit.  This is similar to the soft
behavior of massive gauge field propagator discussed in Appendix A and B,
and to ``softly massive gravity'' emerging in higher dimensional 
constructions \cite {GShifman,Porrati}.

Could (\ref {hybrid}) be a consistent model of a massive graviton in 4D?
There still is a long way to go in order to find out  
whether (\ref {hybrid}) is a theoretically sound theory.
There are three major checks one should perform.

(i) The main problem of the F-P gravity stems from the fact that 
the Hamiltonian of the nonlinear theory is unbounded below.
Hence, one should understand whether the same problem is evaded 
by the hybrid model (\ref {hybrid}). We studied this 
question partially and have shown that the unbounded 
terms that appear in the F-P massive gravity do not arise in 
(\ref {hybrid}). To understand this we look at  the ADM
decomposition of the metric
\begin{displaymath}
\begin{array}{cc}
g_{\mu\nu}=\left(
\begin{array}{cc}
N^2-{\tilde \gamma}_{ij}N^i N^j & -N^j {\tilde \gamma}_{ij} \\
-N^j {\tilde \gamma}_{ij} 	& -{\tilde \gamma}_{ij}
\end{array}
\right),&
g^{\mu\nu}=\left(
\begin{array}{cc}
\frac{1}{N^2}	& -\frac{N^i}{N^2}\\
-\frac{N^j}{N^2} & -{\tilde \gamma}^{ij}+\frac{N^i N^j}{N^2}
\end{array} \right).
\end{array}
\end{displaymath}	
In terms of the new variables the Einstein-Hilbert Lagrangian 
takes the form
\begin{equation}
-{\sqrt{\tilde \gamma}\over 2} N (R^{(3)}+K_{ij}K^{ij}-K^2)\,,
\label{ehnew}
\end{equation}
where ${\tilde \gamma}=\det{{\tilde \gamma}_{ij}}$, $R^{(3)}$ is the 
3-dimension Ricci curvature calculated with the metric ${\tilde \gamma}_{ij}$ 
and the extrinsic curvature tensor
\begin{displaymath}
K_{ij}=\frac{1}{2N}({\dot {\tilde \gamma}}_{ij}-D_i N_j -D_j N_i),
\end{displaymath}
contains a covariant derivative $D_i$ with respect to the metric 
${\tilde \gamma}_{ij}$.
The problem of the F-P gravity arises because the lapse $N$ 
acquires a quadratic term in non-linear realizations of F-P gravity. 
Hence,  it ceases to be a Lagrange multiplier and 
does not restrict the propagation of an extra sixth
``degree of freedom'' which is ghost-like. 
In the hybrid model, however, because of  $\sqrt{g}=1$ we get  
that  $N=1/\sqrt{ |det {\tilde \gamma}_{\mu\nu}|}$. 
This constraint enables one to remove the sixth degree of freedom and  
the dangerous terms that previously led to unboundedness of the 
F-P Hamiltonian \cite {BD}. Regretfully, the expression for the Hamiltonian
of the hybrid model is rather complicated and  
it is difficult to see that there are no other sources 
of rapid classical instability there.

(ii) Even if the rapid  classical instabilities are 
removed, the main question is whether the  BRST invariant 
completion of (\ref {hybrid})
exists. The BRST symmetry, could be a guiding principle 
determining a unique nonlinear completion of 
(\ref {hybrid})(or any other massive theory), 
which at this stage is completely arbitrary.
One could hope that the BRST and ghost charges can be used to 
define the Hilbert space of physical states of the theory 
so that even if the Hamiltonian is 
not bounded below, the states of negative energy do not 
appear in the final states (i.e., they are projected out by 
the conditions $Q_{\rm BRST}|{\rm Phys}\rangle =Q_{\rm Ghost}
|{\rm Phys}\rangle=0$ and cannot be emitted in any process). 
At the moment it is not clear whether such a construction is possible,
but we plan to return to this set of questions in future.

(iii) The question of radiative stability of (\ref {hybrid})
is something one should worry about. In general the Lagrange multipliers 
of the massive theory will acquire the mass and kinetic terms, and this 
would lead to propagation of a new  degree of freedom. 
Only hope here could  be to complete (\ref {hybrid}) in a BRST 
invariant  way so that the resulting theory does not generate 
the quadratic and higher terms for the Lagrange multipliers.

Even if all the above three issues  (i-iii) are positively resolved,
one needs to amend  the massive model to make it consistent 
with the data. The point is that the scalar polarization of a massive graviton
couples to sources and gives rise to contradictions with the  Solar system 
data. In the model of Ref. \cite {DGP} the similar problems is solved 
due to nonlinear effects that  screen the 
undesirable scalar polarization at observable distances 
\cite {ADDG,Gruzinov,Lue,Iglesias}. In the present model, 
at least naively, such a mechanism does not seem to be  
operative, and  some new ideas are needed.

\vspace{0.5in}
{\bf Acknowledgments}
\vspace{0.2in}

We thank Savas Dimopoulos, Gia Dvali, Andrei Gruzinov, 
Nemanja Kaloper, Matthew Kleban, Massimo Porrati,  
Adam Schwimmer, Tom Taylor, Jay Wacker and Dan Zwanziger  
for useful discussions and comments. The work was  supported by 
NASA grant NNG05GH34G. GG is also partly supported by 
NSF grant PHY-0403005, and YS by graduate student 
funds provided by New York University.  YS would also like  to 
thank  Andrei Gruzinov for his support through David and  
Lucile Packard Foundation Fellowship.

\vspace{0.7in}

{\large \bf Appendix A: Qantization of CQED: St\"uckelberg formalism}
\vspace{0.2in}

There is another way of quantizing (\ref {qed1}).
We can restore the gauge invariance of (\ref {qed1}) 
using  the St\"uckelberg method and then 
follow the standard FP procedure of  fixing the gauge 
and introducing  the FP ghosts. Let us discuss this in 
some detail. 

We  start by rewriting  (\ref {qed1})  as follows:
\begin{equation}
\mathcal{L}=-\frac{1}{4}F_{\mu\nu}^2+\lambda(\partial^\mu B_\mu
-\square \varphi)+J^\mu(B_\mu-\partial_\mu\varphi),
\label{stu}
\end{equation}
here we have introduced the notations:
\begin{equation}
B_\mu=A_\mu+\partial_\mu\varphi \quad \textrm{and}\quad
F_{\mu\nu}\equiv\partial_\mu B_\nu
-\partial_\nu B_\mu
=\partial_\mu A_\nu -\partial_\nu A_\mu.
\end{equation}
A Lagrangian, similar to (\ref {stu}), 
in a context of a theory with a non-conserved current was 
recently discussed in \cite{GiaNon}. The Lagrangian (\ref {stu})
is invariant under the 
following gauge transformation
\begin{equation}
\delta B_\mu=\partial_\mu \alpha \quad \textrm{and}\quad \delta\varphi=\alpha,
\end{equation}
where $\alpha$ is an arbitrary function.
To fix  this freedom  we choose a gauge similar to the $R_\xi$-gauge 
that eliminates a mixing terms between $\lambda$ and $B$. 
This can be achieved by adding
into the Lagrangian the following gauge fixing term:
\begin{equation}
\Delta \mathcal{L}_{GF}=\frac{1}{2\xi}(\partial^\mu B_\mu -\xi \lambda)^2.
\end{equation}
Here, $\xi$ is an arbitrary gauge parameter. 
Furthermore, it is easy to find the FP determinant and introduce the FP 
ghosts into the theory. The total Lagrangian that
includes the gauge fixing and FP ghost terms reads:
\begin{equation}
\mathcal{L}_{\rm tot}=-\frac{1}{4}F_{\mu\nu}^2
+\frac{1}{2\xi}(\partial^\mu B_\mu)^2
 - i  {{\bar c}} \square c
-\lambda\square\varphi
+\frac{\xi}{2}\lambda^2+ 
J^\mu(B_\mu- \partial_\mu\varphi).
\label{Lstu1}
\end{equation}
The first three terms on the right hand side of the above
expression constitute a free Lagrangian of gauge-fixed  
QED. In addition there are other states in (\ref {Lstu1}).
To understand their nature, we integrate out from 
(\ref {Lstu1}) the auxiliary field $\lambda $. As a result,
the terms of (\ref {Lstu1}) containing $\lambda$ 
get replaced as
\beq
-\lambda\square\varphi
+\frac{\xi}{2}\lambda^2 \to -\frac{1}{2\xi}(\square\varphi)^2\,.
\label{repl}
\eeq
Because the propagator of 
$\varphi$ is ${\xi\over \square^2} = lim_{\gamma \to 0}
{1\over \gamma} ({\xi \over \square} - {\xi \over \square+\gamma})$,  
it is more appropriate to think of two states, described by  (\ref {repl}), 
one of which has a positive-sign  kinetic 
term and the other one is ghost-like
\footnote{This could  also be understood directly from (\ref {Lstu1}) by
making substitutions: $\lambda=a+b,~\varphi=a-b$,
which generate two kinetic terms one for  $a$ with a 
right sign and another one for  $b$ with a wrong sign. 
These two fields will also have $\xi$-dependent masses and mass mixing
that originate from the $\xi\lambda^2$ term.}.
However, these field are  not present in the physical 
on-shell spectrum of the theory. In the limit 
$\xi \to 0$ the field $\varphi$ is frozen,  
the Lorentz-gauge fixing condition, $\partial^\mu B_\mu =0$, is 
enforced, and the on-shell spectrum consists of two physical
polarizations of a photon. Since physics cannot depend on a choice
of the value of the gauge parameter $\xi$, 
the model (\ref {Lstu1}) propagates  two on-shell 
polarizations. Off-shell, however, there are four 
propagating degrees of freedom. The two non-physical  degrees of freedom 
are longitudinal and time-like components of a photon, while 
$\lambda$ plays a role of the canonically conjugate 
momentum to $A_0$ (alternatively, if $\lambda$ is 
adopted as a canonical coordinate $A_0$ becomes 
its conjugate momentum).
 
It is straightforward to read off the propagators 
from (\ref {Lstu1}) 
\begin{equation}
\Delta^{\mu\nu}_B=-\frac{\eta^{\mu\nu}-(1+\xi)\partial^\mu\partial^\nu/\square}
\square \quad ,~~~ \quad \Delta_\varphi 
\partial^\mu\partial^\nu= -\frac{\xi\partial^\mu\partial^\nu}{\square^2}\,.
\end{equation}
Adding  these two contributions together we
find that the dependence on  the gauge parameter 
$\xi$ cancels out and we obtain
\eqref{eq:propagator_CQED}. This confirms 
our previous conclusion that the $\varphi$ field is 
a gauge artifact. It is also instructive to rewrite the 
Lagrangian (\ref {Lstu1}) in 
terms of the original field $A_\mu$. The latter looks as follows:
\beq
\mathcal{L}_{\rm tot}=-\frac{1}{4}F_{\mu\nu}^2
+\frac{1}{2\xi}(\partial^\mu A_\mu)^2  - {1\over \xi}(\partial^\mu A_\mu)
\square \varphi  -i {{\bar c}} \square c+ A_\mu J^\mu\,,
\label{Lstu2}
\eeq
where we have integrated out the $\lambda$ field.
Here we see a difference from conventional 
QED.  The additional $\xi$ dependent term ensures that 
the resulting free propagator,  for arbitrary values of 
$\xi$, coincides with the Landau gauge propagator of QED. 
This is due to  an extra field $\varphi$
which has no kinetic term but acquires one through the kinetic  
mixing with  the gauge field (\ref {Lstu2}). The 
mixing  term  itself is gauge-parameter dependent
and this is what cancels the $\xi$ dependence of the 
QED propagator.

\subsection*{A.1. On massive deformation of CQED}

In this subsection we study  the massive deformation of the Lagrangian
(\ref {qed1}). We expect, because of the classical constraint,  
the massive theory to be different off-shell from the conventional massive 
electrodynamics (the Proca theory). 

To proceed, we add to the Lagrangian (\ref {qed1}) the following 
mass term:
\beq
\label{massA}
\Delta \l = {1\over 2} m_\gamma^2 A_\mu^2\,. 
\eeq
One can think of this term as arising from some 
higher-dimensional operator in which certain fields acquiring 
VEV's generate (\ref {massA}) while these fields themselves
become heavy and decouple from the low-energy theory.

It is straightforward  to see that Eq. (\ref {eq1}) gets modified
by the mass term
\begin{equation}
\label{eq:EOM_massive_CQED}
\partial^\nu F_{\nu\mu}+m_\gamma A_\mu +J_\mu-\partial_\mu\lambda=0\,,
\end{equation}
while the constraint $\partial_\mu A^\mu=0$ remains intact. 
Taking a derivative
of \eqref{eq:EOM_massive_CQED} one obtains the equation of
motion for $\lambda$ 
\begin{equation}
\square\, \lambda =\partial^\mu J_\mu.
\end{equation}
From the above  we find a solution with a nonzero constant 
background current  $\partial_\mu \lambda=C_\mu$ 
which is identical to that of the constrained massless theory.

It is interesting that the constrained massive theory also has 
a continuous BRST invariance if the FP ghost fields are introduced.
Indeed, consider the Lagrangian
\beq
\l=-\frac{1}{4}F_{\mu\nu}F^{\mu\nu}+\, \lambda (\partial_\mu A^{\mu} )\,
+ {1\over 2} m_\gamma^2 A_\mu^2 \,-\,i {{\bar c}} \square {c}.
\label{mqed1ghost}
\eeq
As before, the presence of the FP ghost in (\ref {mqed1ghost}) does not affect
the discussions of the  classical equations. On the other hand, 
due to these fields the Lagrangian  (\ref {mqed1ghost})
becomes  invariant under the following continuous  BRST transformations:
\beq
\delta A_\mu = i \zeta \partial_\mu c\,,~~~
\delta c = 0,~~~\delta {\bar c} =  \lambda {\zeta}\,,~~~\delta\lambda= 
i m_{\gamma}^2 \zeta c\,,
\label{mBRST}
\eeq
where $\zeta$ is a  continuous Grassmann parameter.
Notice that $\lambda$ transforms to compensate 
for the non-invariance of the mass term\footnote{The above 
transformations (\ref {mBRST}) differ from the standard BRST transformations
of a massive Abelian theory in which massive FP ghost need to be  
introduced. We emphasize here the transformations (\ref {mBRST})
because they will be straightforwardly generalized to the 
non-Abelian massive case in the next section.}. Moreover, 
$\delta^2 A_\mu = \delta^2 c = \delta^2 \lambda =0$, 
while $\delta^2 {\bar c}\neq 0 $, but $\delta^3{\bar c}= 0$.

Let us now turn to the discussion of the spectrum of this theory.
First we evaluate the propagator of the gauge field.
For this we follow the method used in 
section 2. The result is:
\begin{equation}
\Delta^{\mu\nu}=-\lim_{\gamma \rightarrow 0}\left[
{\eta^{\mu\nu}}\,
- \frac{1+\gamma}{\square-\gamma m_\gamma^2}
\partial^\mu \partial^\nu \right]\,
{1\over \square+m_\gamma^2 }\, = 
-\frac {\eta^{\mu\nu}-\partial^\mu\partial^\nu/\square}
{\square+m_\gamma^2} \,.
\label{prop}
\end{equation}
This should be contrasted with  the propagator of
conventional massive QED  (the Proca theory):
\begin{equation}
\Delta_{\rm Proca}^{\mu\nu}
=-\frac{\eta^{\mu\nu}+\partial^\mu\partial^\nu/m_\gamma^2}
{\square+m_\gamma^2}.
\label{propP}
\end{equation}
The key difference of (\ref {prop}) from (\ref {propP})
is the absence in (\ref {prop}) of the longitudinal term that is 
inversely proportional to the mass square. Because of this,
the propagator (\ref {prop}) does have a  good UV
behavior, while  the propagator  (\ref {propP}) does not. In 
an Abelian  theory with a conserved current this hardly matters since
the longitudinal parts of the propagators do not contribute to 
physical amplitudes. However, this could become important 
for  non-Abelian and gravitational theories where  
the matter currents  are only covariantly conserved,
so we carry on with this discussion. 

A natural question that arises 
is what is the mechanism that softens  
the UV behavior of (\ref {prop}) as compared to 
(\ref {propP})? Usually in massive gauge theories this 
is achieved by introducing a Higgs field that regulates the UV
behavior of (\ref {propP}). Therefore,
on top of the three physical polarizations of a
massive gauge field, there should be a new state that 
replaces the role of the Higgs. To see this state manifestly 
we rewrite (\ref {prop}) as follows:
\begin{equation}
\Delta^{\mu\nu}=-\frac{\eta^{\mu\nu}+\partial^\mu\partial^\nu/m_\gamma^2}
{\square+m_\gamma^2}+\frac{\partial^\mu\partial^\nu}{m_\gamma^2\square}\,.
\end{equation}
The first term on the r.h.s. is just the Proca propagator
of three massive polarizations;  the additional term
represents a new massless derivatively-coupled degree of freedom. To uncover 
the nature of this extra state we 
rewrite the Lagrangian of the constrained massive theory as follows:
\begin{equation}
\l=-\frac{1}{4} F_{\mu\nu}^2+{1\over 2} m_\gamma^2 
\left(A_\mu-\frac{\partial_\mu\lambda}{m_\gamma^2}\right)^2
-\frac{(\partial_\mu\lambda)^2}{2 m_\gamma^2}-
i {{\bar c}} \square {c}+ J_\mu A^\mu\,.
\end{equation}
Notice that the quantity  in the parenthesis 
is invariant under the BRST transformations (\ref {mBRST}),
while the non-invariance of the $\lambda$-kinetic term 
under (\ref {mBRST}) is compensated by the terms coming from the 
FP ghost kinetic term.  Defining  a  field 
\begin{equation}
B_\mu\equiv A_\mu-\frac{\partial_\mu\lambda}{m_\gamma^2}\,,
\end{equation}
we end up with the following theory:
\begin{equation}
\label{eq:lagrangian2_massive_CQED}
\l=-\frac{1}{4} F_{\mu\nu}^2+{1\over 2}m_\gamma^2 B_\mu^2+
B_\mu J^\mu-\frac{1}{2m_\gamma^2}
(\partial_\mu\lambda)^2 - i {{\bar c}} \square {c}+
\frac{\partial_\mu\lambda}{m_\gamma^2} J^\mu.
\end{equation}
The first three terms on the r.h.s. of the above expression
represent the Proca Lagrangian of massive electrodynamics.
There are also additional terms in (\ref {eq:lagrangian2_massive_CQED}). 
These are the $\lambda$ kinetic  term  and  FP ghost kinetic term.
These two terms form a sector of the theory that 
is invariant under the continuous BRST transformations 
$\delta \lambda =im_{\gamma}^2 \zeta c$, $\delta {\bar c} =  \zeta\lambda $  
and $\delta c=0$. This symmetry is an exact one if 
the current $J_\mu$ is either conserved, as it is in an Abelian theory,
or transforms w.r.t. BRST in an appropriate way as it will
for a non-Abelian theory discussed in the next section.
The $\lambda$ kinetic  term in (\ref {eq:lagrangian2_massive_CQED})
has a wrong sign, and this state is  ghost-like.
While $J_\mu$ is conserved,  $\lambda$ is decoupled from 
the rest of the physics and can be ignored for all the practical purposes.
Nevertheless,  it is interesting  to understand whether the 
state $\lambda$ could belong to a Hilbert space of physical 
states. This space could  be defined by 
introducing the BRST and Ghost charges and postulating 
$ Q^{\rm BRST}| {\rm Phys}> =0$
and $ Q^{\rm Ghost}| {\rm Phys}> =0$. 
The above conditions could ensure that quanta of
the $\lambda$ field do not belong to the set  of 
physical {\it in} or {\it out} states of the theory.
However, as we discussed previously, the BRST 
transformations (\ref {mBRST}) are rather peculiar
and the construction of the Hilbert space of physical 
states could  differ from the conventional one. 

Here we discuss the properties of the physical degrees of freedom 
in the Hamiltonian formalism.  Let us ignore
the external current $J_\mu$ for the time being.  The canonical momenta 
conjugate to $A_\mu$ read 
\beq
\pi_\mu \equiv {\partial \l
\over \partial {\dot A}^\mu} = -F_{0\mu}+\lambda \delta_{0\mu}\,.
\label{mom}
\eeq
The key difference of this from QED is that
(\ref {qed1}) contains  time derivative of $A_0$, and,  as a result, 
the primary  constraint of QED, $\pi_0=0$, is replaced by the relation
\beq
\pi_0 =\lambda.
\label{primcon}
\eeq
Thus, $\lambda$ is just a conjugate momentum
for $A^0$. The above expression can be used to 
determine $\lambda$, and if so, it does not constrain $\pi_0$. 
Excluding $\lambda$ by (\ref {primcon}), the extended Hamiltonian 
takes the form 
\beq
{\cal H}= {1\over 2} \pi_j^2 \,+\,{1\over 2} (\epsilon_{ijk}\partial_jA_k)^2
+\pi_0 \partial_j A_j + A_0 \left ( \partial_j \pi_j \right )\,.
\label{ham0}
\eeq
To study this system further we look at the  two equations
governing the time evolutions  of $\pi_0$ and
$A_0$ as 
\beq
\label{pbrackets_qed}
{\dot \pi}_0 \,=\,\{\pi_0\,{\cal H} \}\,,~~~~
{\dot A}_0 \,=\,\{A_0\,{\cal H} \}\,,
\label{timevar}
\eeq
where $\{..\}$ denotes the canonical Poisson brackets. This reduces 
to the following relations
\beq
\label{canonical_momenta}
\dot \pi_0+\partial_i \pi_i=0
\,,~~~~~\partial_\mu \,A^\mu =0\,,
\label{con12}
\eeq
both of which were already given in a different form by equations of motion
derived in section 2.  
Furthermore, requiring  that the time derivatives of these 
two expressions are identically zero 
\beq
\{\dot \pi_0+\partial_i \pi_i, {\cal H}\}=0\,,~~~
\{\partial_\mu A^\mu, {\cal H}\}=0,
\eeq
we find the following  two additional equations
\beq
\square \,\pi_0=0\,,~~~~~\square\, A^0+\partial_i \pi_i=0.
\label{con34}
\eeq
Further time derivatives are identically satisfied.

From the first equation of \eqref{con34}
we find that $\pi_0$ can either be a plane  wave or a trivial harmonic 
function (or a superposition of the two).
The second equation of (\ref {con34}), however, dictates 
that $\pi_0$ can only assume the trivial solutions. To show this suppose 
that the solution for $\pi_0$ is a plane wave. Then,
according to 
\beq
\square A^0=-\partial_i \pi_i=\dot \pi_0,
\eeq
$A^0$ cannot have a nonsingular solution because of the mass-shell condition.  
Therefore,  to avoid non-physical solutions
a trivial solution for $\pi_0$ should be taken.
As a result, equation \eqref{con12} and 
\eqref{con34} together remove $\pi_0$ and
$A^0$ from the list of free variables and impose one condition
on $\pi_i$'s and one  on $A^i$'s, thus, reducing 
the number of physical degrees of freedom to two.
This is consistent with the Nakanishi-Lautrup condition that
prescribes to  physical states to satisfy 
$\lambda^{(-)} |{\rm Phys}\rangle= \pi^{(-)}_0 |{\rm Phys}\rangle =0$.

\vspace{0.1in}

In the massive case  the Hamiltonian density is modified to be
\beq
\label{eq:H_massive_CQED}
{\cal H}= {1\over 2} \pi_j^2 \,+\,{1\over 2} (\epsilon_{ijk}\partial_jA_k)^2
+\pi_0 \partial_j A_j + A_0 \left ( \partial_j \pi_j \right )
+\frac{1}{2}m_\gamma^2(A_i^2-A_0^2)\,.
\label{ham0}
\eeq
Time derivatives of $\pi_0$ and $A_0$ 
give rise to the following relations
\beq
\label{eq:con1_massive_CQED}
\dot \pi_0+\partial_i \pi_i=m_\gamma^2 A_0
\,,~~~~~\partial_\mu \,A^\mu =0\,.
\eeq
Their time derivatives lead to Eqs. (\ref {con34}).
The above  four equations, however, are no longer enough to 
remove two extra degrees of freedom. Indeed, 
\beq
\square\, A^0+m_\gamma^2 A^0=-\partial_i \pi_i=\dot \pi_0, 
\eeq
and since  $m_\gamma\neq 0$,  even if $\pi_0$ were a plane 
wave solution of $\square\, \pi_0=0$, non-singular solutions $A^0$
can be obtained.  Therefore $\partial_i \pi_i$ is no longer constrained
to zero, as it was in the massless case.  Hence, 
such a theory propagates three degrees of freedom.

Furthermore, using the relations  
\eqref{eq:con1_massive_CQED} one 
can rewrite the Hamiltonian as follows
\begin{equation}
{\cal H}=\frac{1}{2}\pi_j^2+\frac{1}{2}(\epsilon_{ijk}\partial_j A_k)^2
+\frac{1}{2}m_\gamma^2 \left(A_i-\frac{\partial_i\pi_0}{m_\gamma^2}\right)^2
-\frac{(\partial_i \pi_0)^2}{2 m_\gamma^2}
-\frac{\dot\pi_0 ^2}{2 m_\gamma^2}+\frac{(\partial_i \pi_i)^2}{2m_\gamma^2}.
\end{equation}
Defining  a  new field
\begin{equation}
B_i=A_i-\frac{\partial_i \pi_0}{m_\gamma^2},
\end{equation}
we find the  Hamiltonian 
\begin{equation}
{\cal H}=\frac{1}{2}\pi_j^2+\frac{(\partial_i \pi_i)^2}{2m_\gamma^2}
+\frac{1}{2}(\epsilon_{ijk}\partial_j B_k)^2
+\frac{1}{2}m_\gamma^2 B_i^2-\frac{1}{2m_\gamma^2}
[\pi_\lambda^2+(\partial_i\lambda)^2],
\end{equation}
where the pairs of canonical  coordinates and their conjugate momenta are
$B_i$ and $\pi_i$ and $\lambda$ and $\pi_\lambda$.
The $\lambda$ field makes a negative contribution to
the energy density. However, this field is 
decoupled from all the sources and, thus, cannot be 
produced to grow the negative energy.

\newpage

%\vspace{0.5in}

{\large \bf Appendix B: Constrained non-Abelian gauge fields }
\vspace{0.2in}

Here we generalize  the constrained approach 
to the massless and massive non-Abelian fields.
To understand the essence of these models we 
start with the discussion of the massless case.
The Lagrangian  reads as follows:
\beq
\l=-\frac{1}{4}F^a_{\mu\nu}F^{a\,\mu\nu}+ A^a_\mu J^{a\mu}+ 
\lambda^a (\partial_\mu A^{\mu} )^a\,.
\label{qcd1}
\eeq
Equations of motion for the gauge fields and Lagrange multiplier are 
respectively:
\beq
D^\nu F^a_{\nu\mu}+ J^a_\mu -\partial_\mu \lambda^a =0 \,, 
\label{qcdeq1}\\
\partial_\mu A^{a\mu}\,=\,0\,.
\label{qcdeq2}
\eeq
An equation for $\lambda$  follows by taking a covariant derivative of 
(\ref {qcdeq1})
\beq
D^\mu\partial_\mu \lambda^a = D^\mu J^a_\mu \,.
\label{qcdlJ}
\eeq
In a theory with a covariantly conserved source there could exist 
new non-trivial classical solutions for $\lambda$ such that 
$\partial _\mu \lambda$  is also covariantly conserved.
One particular solution is a path-ordered Wilson line
\beq
\partial _\mu \lambda (x) = {\cal P} {\rm exp} \left (-ig \int _y^x A_\mu(z) 
dz^\mu  \right )\partial _\mu \lambda (y)\,. 
\label{Wilson}
\eeq
Thus, if there is a non-zero external colored current solution
$\partial _\mu \lambda$ at some point, then its value at any other 
point can be calculated according to (\ref {Wilson}).
 
The next issue to be address is that of 
quantum consistency of this approach. Again an important fact is that
the Lagrangian (\ref {qcd1}) can be completed
to a BRST invariant form:
\beq
\l=-\frac{1}{4}F^a_{\mu\nu}F^{a\,\mu\nu}+ A^a_\mu J^{a\mu}+ 
\lambda^a (\partial^\mu A^a_{\mu} )\,- 
i {\bar c}^a \partial^\mu D^{ab}_\mu c^b\,.
\label{qcd2}
\eeq
Since  at the classical solutions that we consider the FP ghost
fields  vanish, (\ref {qcd2}) recovers all the classical 
results of (\ref {qcd1}). The explicit
BRST transformations leaving  (\ref {qcd2}) invariant are:
\beq
\delta A^a_\mu = i\zeta D^{ab}_\mu c^b\,,~~~
\delta c^a = -{i\over 2} g \zeta f^{abd}c^bc^d\,,~~~
\delta {\bar c}^a = \lambda^a {\zeta}\,,~~~\delta\lambda^a=0\,.
\label{BRSTn}
\eeq
The expression for the BRST and Ghost currents, as well as  
the subsidiary conditions that guarantee  the unitarity and completeness
of the physical Hilbert space of states are the standard ones \cite {Kugo}.

As in the case of a photon, the massless free propagator reads:
\beq
\Delta^{ab}_{\mu\nu} = -\delta^{ab}\,
{ \eta_{\mu\nu} -\partial_\mu\partial_\nu/\square 
\over \square}\,. 
\label{freepropnon}
\eeq
Finally, due to the presence of the BRST symmetry 
the Lagrange multiplier does not acquire kinetic term
via the loop corrections. This is because the two-point
correlation functions of the $A_\mu$ field is guarantied to be 
transverse due to the presence of the FP ghost and the 
transverse structure of the tree-level propagator  
(\ref {freepropnon})
\beq
\partial^\mu \lambda^a \, \langle A^a_\mu \,A^b_\nu \rangle \partial ^\nu 
\lambda^b
\sim \partial^\mu \lambda^a \delta^{ab} 
(\eta_{\mu\nu} \square -\partial_\mu \partial_\nu)
\partial ^\nu \lambda^b =0 \,,
\label{gener1}
\eeq 
to all orders of perturbation theory.

\subsection*{B.1. St\"uckelberg formalism}

Instead of using BRST symmetry to arrive from (\ref {qcd1})  
to (\ref {qcd2}), here we follow a conventional  route.
First we restore  gauge symmetry of (\ref {qcd1})
and then fix that restored gauge invariance and introduce 
the appropriate FP ghosts. Let us start by defining 
new variables:
\beq
ig A_\mu = U^+D_\mu U,~~~{\rm where}~~~U=e^{it^a\pi^a},~~~~
D_\mu =\partial_\mu + igB_\mu\,,
\label{qcdB}
\eeq
and rewrite the Lagrangian (\ref {qcd1}) in the following form
\beq
\l= {1\over 2} {\rm Tr} \left (  -\frac{1}{4}F_{\mu\nu}F^{\mu\nu} 
+ {\lambda \partial^\mu [U^+D_\mu U]\over ig} + {U^+D_\mu U\over ig} 
J^\mu \right )\,. 
\label{qcd11}
\eeq
The above Lagrangian is gauge invariant under the local transformations 
of $U\to e^{i\alpha^at^a}U$ and $B_\mu \to e^{i\alpha^at^a}B_\mu 
e^{-i\alpha^at^a}+{i\over g}
[\partial_\mu e^{i\alpha^at^a} ] e^{-i\alpha^at^a}$,
where $t^a$ denote the generators of a local gauge group,
and $\alpha^a$'s are gauge transformation functions\footnote{
Notice that the current $J_\mu$ and the Lagrange multiplier $\lambda$
are not supposed to transform under these gauge transformations.
This can be achieved by rewriting the fundamental fields out of which 
$J_\mu$ is constructed (as well as rewriting $\lambda$) in terms of
new fields rescaled by $U$'s. Under the gauge transformations 
the new field transform in a conventional way but their 
variance is compensated by transformations of $U$'s, 
so that $J_\mu$  and $\lambda$ stay invariant.}.
We can proceed further and  fix this 
gauge freedom by introducing into the Lagrangian the following term 
\beq
{1\over 4\xi} {\rm Tr} ( \partial_\mu B^\mu -\xi \lambda)^2\,. 
\label{gfqcd1}
\eeq
This term removes quadratic mixing between $B_\mu$
and $\lambda$, and   after integrating out
$\lambda$ and rewriting it back in terms of the $A_\mu$ field, 
the result  reads as follows:
\beq
\l= {1\over 2} {\rm Tr} \left (  -\frac{1}{4}F_{\mu\nu}F^{\mu\nu}+
{1\over 2\xi} ( \partial_\mu A^\mu)^2 +{1\over \xi} ( \partial_\mu A^\mu)
(\partial^\mu \Pi_\mu) - i {\bar c}\partial_\mu D_\mu c+     A_\mu J^\mu  
\right ), 
\label{qcd123}
\eeq
where $\Pi_\mu= U A_\mu U^+ +{i\over g} [\partial_\mu U]U^+ -A_\mu$.
All the terms of the Lagrangian (\ref {qcd123}), except the third one, 
is what one gets in the conventional approach. The third, 
gauge dependent term, has a structure that guarantees
that the free propagator for an arbitrary value of $\xi$ 
coincides with the Landau gauge propagator of the  
conventional approach. This is due to the extra fields $\pi^a$'s 
which by themselves have no kinetic term but acquire one
through the gauge-parameter-dependent kinetic 
mixing term  with $A_\mu$ in (\ref {qcd123}).

\subsection*{B.2. Comments on massive constrained non-Abelian fields}

Let us start with a component form of the constrained massive Lagrangian 
\beq
\l=-\frac{1}{4}F^a_{\mu\nu}F^{a\,\mu\nu}+ A^a_\mu J^{a\mu}+ 
\lambda^a (\partial_\mu A^{\mu} )^a\,+{1\over 2} M^2 A^a_\mu A_\mu^a\,-
i {\bar c}^a \partial^\mu D^{ab}_\mu c^b\,.
\label{mqcd}
\eeq
It is interesting that the action of this theory is BRST invariant
under the following transformations 
(the Lagrangian transforms as a total derivative)
\beq
\delta A^a_\mu = i\zeta D^{ab}_\mu c^b\,,~~~
\delta c^a = -{i\over 2} g \zeta f^{abd}c^bc^d\,,~~~
\delta {\bar c}^a =  \lambda^a {\zeta}\,,~~~
\delta\lambda^a= i M^2 \zeta c^a\,.
\label{mBRSTn}
\eeq
Notice that the Lagrange multiplier does transforms w.r.t. the BRST
and compensates for the non-invariance of the mass term. 
The above  BRST transformations have the following peculiar properties:
\beq
\delta^2 A^a_\mu =\delta^2 c^a=0; ~~\delta^2 {\bar c}^a\neq 0\neq
\delta^2 \lambda^a; ~~\delta^4 {\bar c}^a=\delta^3 \lambda^a =0\,.
\label{BRSTpec}
\eeq
The respective BRST and Ghost currents (the latter is due to the 
ghost-rescaling symmetry) ${\bar c}\to e^\alpha {\bar c}, 
~{c}\to e^{-\alpha} {c}$) can be derived
\beq
J_\mu^{\rm BRST} = -F_{\mu\nu} D_\nu c + \lambda D_\mu c -{ig\over 2}
\partial_\mu ({\bar c}^a)f^{abd}c^bc^d,~~~J_\mu^{\rm Ghost} =
i ( {\bar c^a} D_\mu c^a - (\partial_\mu {\bar c}^a) c^a )\,.
\label{JJ}
\eeq
The physical states could be defined in analogy with the standard 
approach
\beq
Q_{\rm BRST} |{\rm Phys}> =0, ~~~~ Q_{\rm Ghost} |{\rm Phys}> =0\,,
\label{Qbrst}
\eeq
where the BRST and Ghost charges are defined as $Q_{\rm BRST} =\int d^3x
J^{\rm BRST}_0(t,x)$ and $Q_{\rm Ghost} =\int d^3x J^{\rm Ghost}_0(t,x)$. 
This suggest that the $\lambda$ state, which upon the 
diagonalization of the Lagrangian acquires a ghost-like kinetic term,
should not be a part of the physical Hilbert space of {\it in} and {\it out} 
states of the theory. The fictitious $\lambda$ particle should be 
allowed to propagate as an intermediate state in Feynman diagrams softening 
the  UV behavior of the theory, however, it cannot be emitted as
a final {\it in} or {\it out} state of the theory.
In this respect it should be  similar to a FP ghost. However,
because the peculiar properties of the above BRST transformations 
(\ref {BRSTpec}), it still remains to be seen 
that for a rigorous construction of a positive semi-definite norm
Hilbert space of states with unitary S-matrix elements
the conditions (\ref {Qbrst}) are enough.
Besides the quantum effects, one should make sure that 
rapid classical instabilities are also removed.
These may require further modification of the model.
Detailed studies of this issue will be presented elsewhere.

Having the formalism of the previous subsection developed
it is easy now to discuss the spectrum of massive theory.
First we restore the gauge invariance of the massive Lagrangian
(\ref {mqcd}) by using the variables (\ref {qcdB}), and then gauge fix
it by (\ref {gfqcd1}). The resulting Lagrangian is
\beq
\l= {1\over 2} {\rm Tr} \left (  -\frac{1}{4}F_{\mu\nu}F^{\mu\nu} 
+ \lambda \partial^\mu \left ({U^+D_\mu U\over ig} \right )+ {M^2\over 2}
\left ({U^+D_\mu U\over ig} \right )^2 
+ {1\over 2\xi} ( \partial_\mu B^\mu -\xi \lambda)^2 \right ). 
\label{qcd12}
\eeq
The mixing term between the gauge field and 
Goldstones is canceled and  we  
integrate out the $\lambda$ field. The resulting Lagrangian reads:  
\beq
\l= {1\over 2} {\rm Tr} \left (  -\frac{1}{4}F_{\mu\nu}F^{\mu\nu}(A) 
+ {1\over 2}M^2 A_\mu^2 + 
{1\over 2\xi} ( \partial_\mu A^\mu)^2 +{1\over \xi} ( \partial_\mu A^\mu)
(\partial^\mu \Pi_\mu) + A_\mu J^\mu  
\right ), 
\label{masqcd1}
\eeq
where, as before, we defined 
$\Pi_\mu= U A_\mu U^+ +{i\over g}[\partial_\mu U]U^+-A_\mu$.
 
A remarkable feature of this model is that 
the propagator takes the form:
\beq
\Delta^{ab}_{\mu\nu} = -\delta^{ab}\,\left (
{ \eta_{\mu\nu} -{(1+\xi)\over \square -\xi M^2} \partial_\mu\partial_\nu} 
\right )\,{1\over \square+M^2}\,- \delta^{ab}{\xi \partial_\mu \partial_\nu
\over (\square -\xi M^2)\square}\,, 
\label{propmqcd}
\eeq
which has a smooth  UV behavior and non-singular massless limit.

%\newpage

\end {document}